\documentclass[aps,amsmath,prapplied,reprint,groupedaddress]{revtex4-1}
\usepackage{graphicx}
\usepackage{dcolumn}
\usepackage{bm}
\usepackage{amsfonts}
\usepackage{array}
\usepackage[colorlinks,linkcolor=red,anchorcolor=blue,citecolor=green]{hyperref}

\begin{document}


\title{Continuous-variable Quantum Key Distribution with Rateless Reconciliation Protocol}


\author{Chao~Zhou$^{1}$, Xiangyu Wang$^{1}$}
\author{Yichen Zhang$^1$}
\email[]{zhangyc@bupt.edu.cn}
\author{Zhiguo Zhang$^1$}
\author{Song Yu$^1$}
\author{Hong Guo$^2$}
\email[]{hongguo@pku.edu.cn}

\affiliation{$^1$State Key Laboratory of Information Photonics and Optical Communications, Beijing University of Posts and Telecommunications, Beijing 100876, China}
\affiliation{$^2$State Key Laboratory of Advanced Optical Communication, Systems and Networks, Department of Electronics, and Center for Quantum Information Technology, Peking University, Beijing 100871, China}


\date{\today}

\begin{abstract}

Information reconciliation is crucial for continuous-variable quantum key distribution (CV-QKD) because its performance affects the secret key rate and maximal secure transmission distance. Fixed-rate error-correction codes limit the potential applications of the CV-QKD because of the difficulty of optimizing such codes for different low SNRs. In this Paper, we propose a rateless reconciliation protocol combined multidimensional scheme with Raptor codes that not only maintains the rateless property but also achieves high efficiency in different SNRs using just one degree distribution. It significantly decreases the complexity of optimization and increases the robustness of the system. Using this protocol, the CV-QKD system can operate with the optimal modulation variance which maximizes the secret key rate. Simulation results show that the proposed protocol can achieve reconciliation efficiency of more than 95\% within the range of SNR from -20 dB to 0 dB. It also shows that we can obtain a high secret key rate at arbitrary distances in a certain range and achieve a secret key rate of about $5\times10^{-4}$ bits/pulse at a maximum distance of 132 km (corresponding SNR is -20dB) that is higher than previous works. The proposed protocol can maintain high efficient key extraction under the wide range of SNRs and paves the way toward the practical application of CV-QKD systems in flexible scenarios.

\end{abstract}


\maketitle

\section{Introduction}
Quantum key distribution (QKD) \cite{gisin2002quantum,scarani2009security,pirandola2019advances} is one of the most practical applications of quantum-information technologies. QKD enables two spatially separated parties named Alice and Bob to share random keys in the untrusted environment and promises unconditional security in principle \cite{diamanti2016practical}. With the development of quantum-computer research, the existing classical encryption methods based on computational complexity are threatened. Under the influence of such threats, QKD based on physical properties has attracted worldwide attention. The quest for high-performance QKD systems in the last few years has led to several successful demonstrations based on different protocols.

There are two types of protocols for generating symmetric keys over quantum channel: discrete-variable QKD (DV-QKD) \cite{gisin2002quantum,scarani2009security,bennett2014quantum} and continuous-variable QKD (CV-QKD) \cite{braunstein2005quantum,weedbrook2012gaussian,diamanti2015distributing}. In DV-QKD, the information is encoded in the polarization of single-photon states and single-photon detector is used to measure the received quantum state. In CV-QKD, the information is encoded in the amplitude and phase quadratures of quantum states and heterodyne or homodyne detection techniques are used in this case.
CV-QKD has attracted much attention as it offers the possibility for implementations based on classical telecom components \cite{grosshans2002continuous,grosshans2003quantum,Lodewyck2007Quantum,PhysRevA.88.010302,jouguet2013experimental,weedbrook2014two,pirandola2015high,soh2015self,qi2015generating,PhysRevA.92.062337,PhysRevA.95.062330,PhysRevApplied.10.064028,PhysRevApplied.9.054008,8304639,8767074,eriksson2019wavelength,Ye:19,PhysRevA.99.032327,PhysRevX.9.021059,Wang:19,zhang2019one}.
For a CV-QKD protocol based on coherent states with Gaussian modulation, a composable security proof against arbitrary attacks has been provided \cite{pirandola2008characterization,pirandola2009direct,PhysRevLett.114.070501,PhysRevLett.118.200501}. Moreover, some experiments based on CV-QKD protocols have been successfully implemented in commercial links and obtained high secret key rate at low repetition rate \cite{Zhang_2019}. The integrated silicon photonic chip for CV-QKD systems promotes real-world applications of on-chip hybrid quantum-classical communication for advanced communication networks\cite{zhang2019integrated}.

A typical CV-QKD system consists of two parts, the physical link and the information postprocessing \cite{milicevic2018quasi,Wang2019}. In the first part, Alice prepares quantum states and sends them to Bob through a quantum channel. Then Bob measures quantum states using a homodyne detector. The second part is the process of dealing with information and getting secret keys operated by both sides.
Information reconciliation in postprocessing is one of main factors limiting the transmission distance of CV-QKD system \cite{jouguet2011long,van2004reconciliation}.
Ref.~\cite{leverrier2008multidimensional} proposed multidimensional reconciliation method which provides a way to use the classical error-correcting codes and improves the performance of the CV-QKD system under low SNR.
Reconciliation efficiency is an important parameter, which is the usual expression of the secret key rate taking into account the imperfect reconciliation protocol \cite{Leverrier2010Finite}.
Ref.~\cite{jouguet2013experimental} used the multiedge-type low-density parity check (MET LDPC) codes in conjunction with multidimensional reconciliation method to achieve high-reconciliation efficiency.

The longer the communication distance, the higher the reconciliation efficiency required to ensure a high secret key rate. It is a real challenge to obtain high secret key rate at such long distance. Because the SNR of the quantum channel may be lower than - 15 dB or even - 20 dB, it is difficult to correct errors under such conditions \cite{chung2001design,richardson2002multi}.
At present, MET LDPC codes are fixed-rate codes which can only obtain high error-correction performance at the corresponding SNRs to these codes \cite{richardson2001design,wang2018high00,milicevic2018quasi}.
However, it is very difficult to design them at low rate with long block length on the order of $10^6$ bits \cite{leverrier2009unconditional}.
In different application circumstances, the SNR of a CV-QKD system is also different. Since the performance of the MET LDPC codes is very sensitive to slight changes in SNRs. The reconciliation efficiency is decreased when the practical SNR differs from the codes' optimal suitable SNR \cite{wang2017efficient}. Thus a finite number of designed codes cannot support fully practical applications. Besides, it is not realistic and over complex to find all available codes for each different practical SNR.
This motivates us in this work to break through these restrictions.

In this Paper, we propose a rateless reconciliation protocol based on Raptor codes \cite{shokrollahi2006raptor}. The rateless codes can generate a potentially limitless number of coded symbols for a given set of information symbols.
Thus the rate of these codes is uncertain before information transmission.
We choose Raptor codes as the error-correcting codes because they are the first rateless codes with linear time encoding and decoding and have been used in several applications with large data transmission.
Raptor codes were studied for additive white Gaussian noise (AWGN) in Refs.~\cite{etesami2006raptor,cheng2009design,kuo2014design,shirvanimoghaddam2016raptor}, which proposed a method to find the optimal degree distribution in a given SNR.
The rateless property of these codes makes them easier to be optimized for different SNRs. Compared with fixed-rate codes, the design complexity of rateless codes is reduced.
This protocol can maintain the property of rateless codes and use just one degree distribution to achieve high reconciliation efficiency under a wide range of SNRs.
Additionally, in previous CV-QKD systems, the modulation variance is adjusted in real time so as to be as close as possible to the SNR corresponding to the threshold of an available fixed-rate code \cite{jouguet2013experimental}. Although this approach can achieve high reconciliation efficiency, it also sacrifices the optimal modulation variance. What is more, it is hard to reach the expected accurate SNRs.
The rateless reconciliation protocol allows the modulation variance to maintain an optimal value, which can improve the performance of the system. It is suitable to CV-QKD systems in different scenarios.

This Paper is organized as follows: in Sec.~\ref{sec2} a brief review of postprocessing in CV-QKD systems is given. In Sec.~\ref{sec3}, we describe the details of the rateless reconciliation protocol. In Sec.~\ref{sec4}, several simulation results are carried out to fully evaluate these advantages. Finally, we conclude this Paper with a discussion in Sec.~\ref{sec5}.

\section{postprocessing in CV-QKD}
\label{sec2}
A CV-QKD system consists of two legitimate parties, Alice and Bob. Alice prepares Gaussian-modulated coherent states and sends them to Bob who measures one of the quadratures with homodyne detection.
After Bob measures quantum states sent from Alice through the quantum channel, both sides start the postprocessing to extract the secret keys over an authenticated classical public channel, which is assumed to be noiseless and error free. In this Paper, we use a reverse reconciliation scheme in which Alice and Bob use Bob's data to obtain the secret key.
\subsection{Postprocessing procedure}
The postprocessing of a CV-QKD system contains four steps: base sifting, parameter estimation \cite{Leverrier2010Finite,jouguet2012analysis}, information reconciliation \cite{van2004reconciliation,leverrier2008multidimensional,jouguet2011long,grosshans2003virtual,jiang2017secret} and privacy amplification \cite{bennett1995generalized,deutsch1996quantum,wang2018high}. Base sifting refers to Bob sending a randomly selected measurement base to Alice. Then Alice keeps the correlated raw data according to the measurement bases. The purposes of the parameter estimation procedure are to determine quantum-channel parameters and estimate the secret key rate. During information reconciliation, Alice and Bob can extract available common sequences from their correlated raw data. After the above steps, Eve may have collected sufficient information during her observations of the quantum and classical channels. Hence, privacy amplification is an indispensable step, which is used to distill the final secret keys from the common sequence between Alice and Bob.

Let us discuss the details of the parameter estimation step, which is relevant to calculate the secret key rate. According to the Gaussian optimality theorem, Alice and Bob's two-mode state at the output of the quantum channel is fully characterized by Alice's modulation variance $V_{A}$, channel transmission $T$ and excess noise $\xi$, which is added by the channel. With these notations, all noises are expressed in shot noise units.
In order to calibrate the shot-noise, it is necessary to obtain the electric noise $\nu_{\mathrm{el}}$ and the efficiency of the homodyne detection $\eta$ firstly. These two parameters are assumed not to be accessible to Eve and are measured with a large amount of data during a secure calibration procedure that takes place before the deployment of the system. The parameters $V_{A}$, $T$ and $\xi$ are estimated in real time by using a fraction of raw data after base sifting. Here we assume the standard loss of a single-mode optical fiber cable to be $\alpha=0.2dB/km$.
For a CV-QKD protocol, the modulation variance $V_{A}$ is one of the main physical parameters that influence the secret key rate. Therefore, $V_{A}$ should be kept at the optimal value to maximize the expected secret key rate.

Taking finite-size effects into account, the secret key rate of a CV-QKD system with one-way reverse reconciliation is given by \cite{Leverrier2010Finite}:
\begin{equation}\label{finite_secret_key}
K_{\mathrm{finite}}=\frac{n}{N}\left [ \beta I(A:B)-S_{\epsilon_{\mathrm{PE}}}(B:E)-\bigtriangleup(n)\right ],
\end{equation}
where $N$ is the total number of data exchanged by Alice and Bob, $n$ is the number of data used for key extraction, and the other $m=N-n$ data is used for parameter estimation. $\beta \in {[0,1]}$ is the reconciliation efficiency, and $I(A:B)$ is the classical mutual information between Alice and Bob. $S_{\epsilon_{\mathrm{PE}}}(B:E)$ is the maximum of the Holevo information that Eve can obtain from the information of Bob, where $\epsilon_{\mathrm{PE}}$ is the failure probability of parameter estimation. $\bigtriangleup(n)$ is the finite-size offset factor. Thus an imperfect reconciliation scheme results in the reduction of the secret key rate and limitation of the range of the protocol. In order to achieve a high key rate, a higher reconciliation efficiency $\beta$ is needed under the condition of low SNRs.

\subsection{Raptor codes}
\begin{figure}
	\centering
	\includegraphics[width=1.0\linewidth]{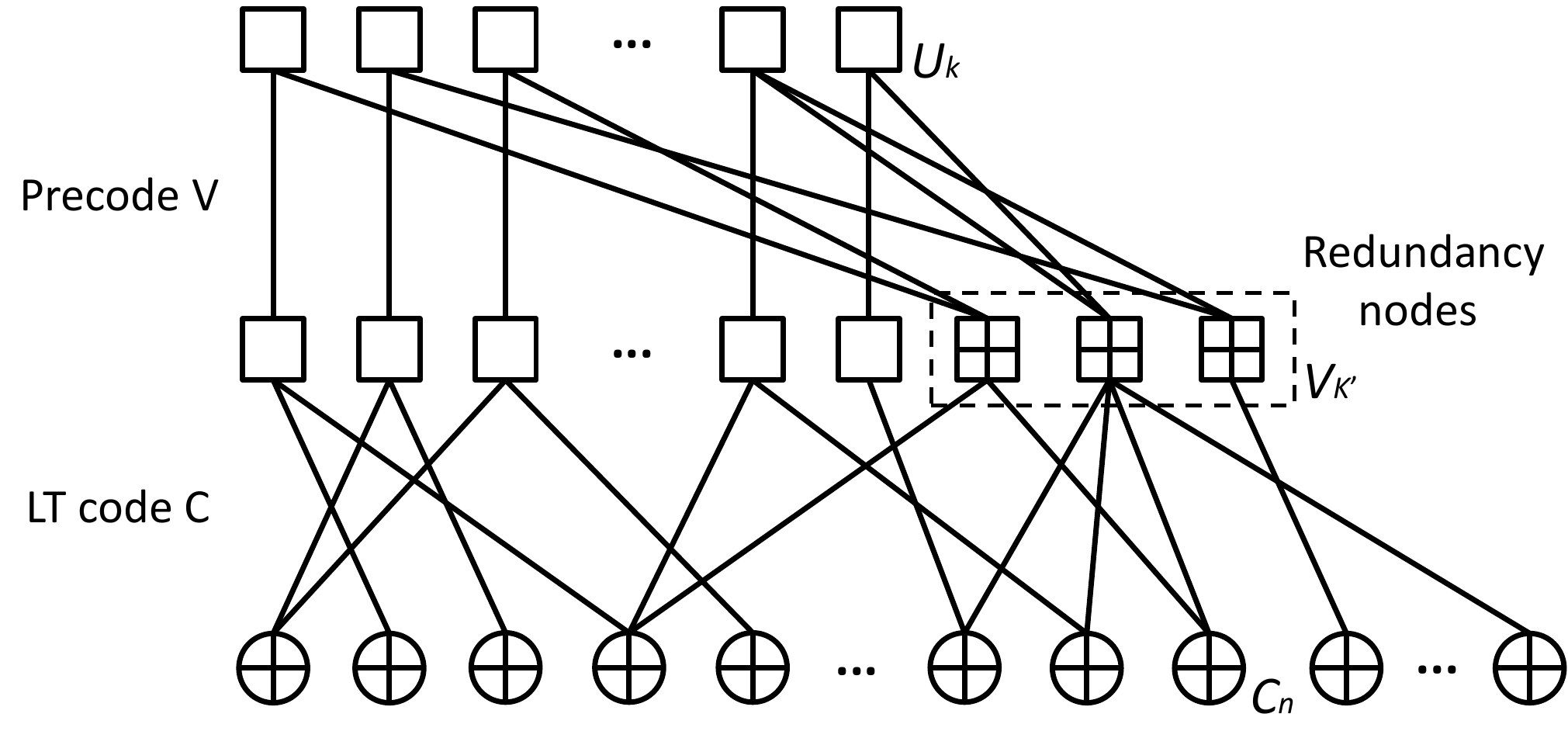}
	\caption{Factor graph for a Raptor code. $U_{k}$ denotes the initial $k$ bits. $V_{k'}$ denotes $k'$ input bits of $C$. $C_{n}$ denotes the output $n$ bits. The nodes in the dashed border are redundancy nodes generated by precoding.}
	\label{fig:raptor-code}
\end{figure}

Error correction is a part of information reconciliation that affects the reconciliation efficiency. In this Paper we introduce Raptor codes as the error-correction codes for CV-QKD. Here we give some basics of these codes.
Figure~\ref{fig:raptor-code} shows the factor graph for a Raptor code. In general, a Raptor code includes two parts: linear precoder $V$ and LT code $C$.
A Raptor code can be characterized by ($k$, $V$, $\Omega(x)$), where $\Omega(x):=\sum_{d}\Omega_{d}x^{d}$ is the degree distribution polynomial and $\Omega_{d}$ denotes the probability of an output node with degree $d$. In this Paper, we choose the LDPC code as the precoding code. In general, the encoding process of the Raptor code is as follows:
\begin{enumerate}
	\item A high-rate check matrix is selected and converted into a generator matrix, which is used to encode the initial $k$ bits. Through the LDPC encoding, $k'$ bits is generated.
	\item From a given degree distribution $\Omega(x)$, randomly select a degree $\omega_{i}\left \{ 0<i<n \right \}$ for the output bit $c_{i}$.
	\item Choose $\omega_{i}$ distinct message bits, uniformly at random. Use an index set $G_{i}$ to denote which message bits are selected.
	\item Calculate the final value of the output bit $c_{i}$ by XOR operations of $\omega_{i}$ message bits, i.e., $c_{i}=v_{g_{i},1}\bigoplus v_{g_{i},2}\bigoplus ...\bigoplus v_{g_{i},\omega_{i}}$.
	\item Repeat the above steps 2)-4) until enough output bits are generated.
\end{enumerate}

For the convenience of analysis, the output bits $c_{1},c_{2},...,c_{n}$ are sent to the receiver through a given binary input AWGN (BIAWGN) channel. After enough bits $y_{1},y_{2},...,y_{n}$ have been received, the receiver starts decoding, where $y_{j}=(-1)^{c_{j}}+z_{j}$ and $z_{j}$ is the zero mean Gaussian noise with variance $\sigma_{c}^{2}$ for every $j\in {1,2,...,n}$. The corresponding channel SNR is defined as $\gamma_{c}=1/\sigma_{c}^{2}$. On the factor-graph representation of Raptor codes in Fig.~\ref{fig:raptor-code}, the sum-product algorithm can be applied to decode $C$. The channel log likelihood ratio (LLR) message of $c_{j}$ is defined as:
\begin{equation}\label{LLR_message_of_cj}
m_{j}^{0}:=\mathrm{log}\frac{P(c_{j}=0|y_{j})}{P(c_{j}=1|y_{j})}.
\end{equation}

The sum-product algorithm operates in an iterative way where messages are passed bidirectionally along each edge in the factor graph between neighboring input bits and output bits. At the $p$th iteration, we denote the message passed from output node $o$ to input node $i$ by $m_{o\rightarrow i}^{(p)}$ and the message passed from input node $i$ to output node $o$ by $m_{i\rightarrow o}^{(p)}$. In each iteration, every node passes messages to its neighbors along its edges. Then the message passed from output node to input node at each iteration $p$ are formulated as follows\cite{kuo2014design}:
\begin{equation}\label{message_updating1_for_decoding}
m_{o\rightarrow i}^{(p)}=2\mathrm{tanh}^{-1}[\mathrm{tanh}(\frac{m_{o}^{(0)}}{2})\prod_{i'\neq i}\mathrm{tanh}(\frac{m_{i'\rightarrow o}^{(p-1)}}{2})],
\end{equation}
and the message passed from input node to output node is
\begin{equation}\label{message_updating2_for_decoding}
m_{i\rightarrow o}^{(p)}=\sum_{o' \neq o}m_{o' \rightarrow i'}^{(p)}.
\end{equation}

After a predetermined maximum number of iterations $P$, the decoded LLR, $m_{i}$, for input node $i$ is computed as
\begin{equation}\label{message_updating3_for_decoding}
m_{i}=\sum_{o}m_{o \rightarrow i}^{(P)}.
\end{equation}

The decoded LLRs of the input nodes are passed to code $V$ to recover the original message bits. The entire decoding process is repeated as gradually increasing the number of output bits until $u_{1},u_{2},...,u_{k}$ are correctly decoded. Generally, the decoder uses the check equations of precoder $V$ to verify the correction of the decoding results. If this round of decoding fails, the receiver collects more bits from the sender and starts to decode again. Once the decoding is successful, the receiver sends a stop signal to the sender through a feedback channel.

In the next section, we introduce the rateless reconciliation protocol based on Raptor codes.
The multidimensional reconciliation method transforms a channel with a Gaussian modulation to a virtual binary modulation channel, with a capacity loss that is very low at low SNR. This method enables the Raptor codes to be applied in the BIAWGN channel. The proposed protocol not only maintains the property of rateless codes but also achieves high reconciliation efficiency in different SNRs.

\section{Rateless reconciliation protocol}
\label{sec3}
\begin{figure}
	\centering
	\includegraphics[width=0.85\linewidth]{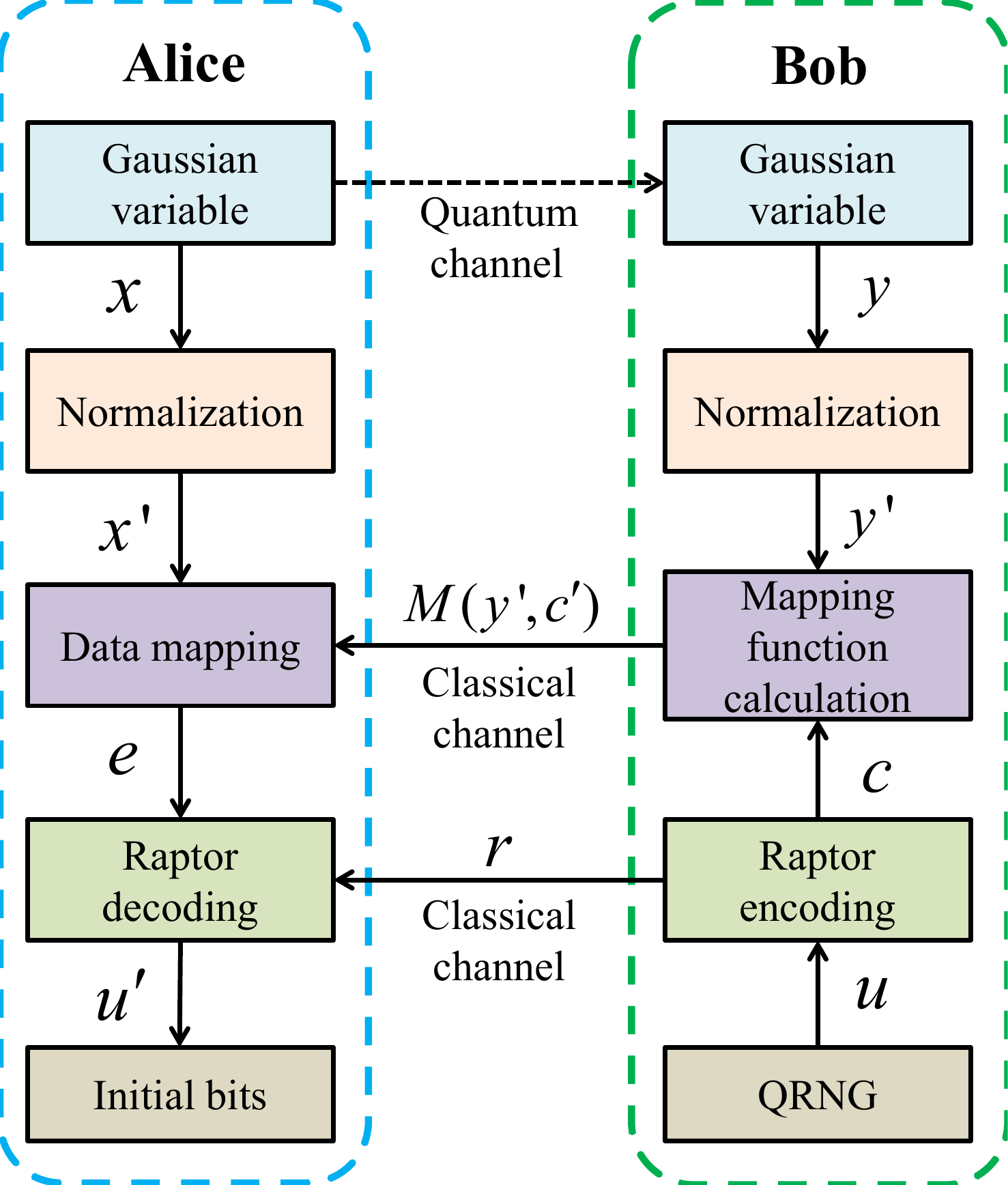}
	\caption{Schematic diagram of rateless reconciliation protocol. $x$ and $y$ are two correlated Gaussian sequences. $x'$ and $y'$ represent the normalized sequences. $M(y',c')$ represents the mapping function sent from Bob to Alice. $u$ denotes the initial sequence generated by QRNG. $c$ is the Raptor encoding result of $u$. $e$ is the sequence before decoding and $u'$ is the decoding result that is equal to $u$ when the decoding is successful. $r$ denotes the additional check codes. QRNG, quantum random-number generator.}
	\label{fig:reconciliation-protocol}
\end{figure}

The rateless reconciliation protocol of CV-QKD can be divided into two parts. First, Alice and Bob use the multidimensional reconciliation method to generate discrete variables by rotate Gaussian variables. Then, Alice and Bob correct all errors between their sequences using Raptor codes. The schematic diagram of the proposed protocol is shown in Fig.~\ref{fig:reconciliation-protocol}.

According to this schematic diagram, $x$ and $y$ are two correlated Gaussian sequence, satisfying $x\sim N(0,\sum^2)$, and the property satisfies $y=x+z$, $z\sim N(0,\sigma^2_Z) $. Alice and Bob choose $d$ to divide these sequences, where $d$ is the dimension of multidimensional reconciliation. Practical CV-QKD systems mainly adopt eight-dimensional reconciliation ($d=8$) because it has the highest performance compared with other dimensions ($d=1, 2, 4$) \cite{leverrier2008multidimensional}. Then Alice and Bob normalize their Gaussian variables $x$ and $y$ to $x'$ and $y'$, respectively, with $x'=x/\parallel x \parallel$, $y'=y/\parallel y \parallel$, where $\parallel x \parallel=\sqrt{ \langle x,x \rangle}$, $\parallel y \parallel=\sqrt{ \langle y,y \rangle}$. Sequences $x'$ and $y'$ have a uniform distribution on the unit sphere $S^{n-1}$ of $\mathbb{R}^{n-1}$.
The binary sequence $u$ is generated by quantum random-number generator and follows uniform distribution for the security of the CV-QKD system.
In Fig.~\ref{fig:reconciliation-protocol}, we can see that $c$ is generated by random binary sequence $u$ through Raptor encoding. The detailed process of Raptor encoding is studied in section \ref{sec2}.

It is worth mentioning that the coding characteristic of the LT code is to select message bits randomly according to a degree distribution and generate limitless output bits. The uniformity of the probability distribution of Bob's variables on $\mathbb{F}_{2}^{n}=\left \{ 0,1 \right \}^{n}$ is an essential assumption in order to prove that the side information Bob sends to Alice on the public channel does not give any relevant information to Eve about the code word chosen by Bob. The binary sequence $c$ having a uniform distribution is a necessary and sufficient condition for the multidimensional reconciliation method.
This condition is guaranteed by the encoding process of the Raptor code.
Let index set $g_{i}$ denote which message bits are selected, $v$ denotes the input bits of the LT code, and $\omega_{i}$ denotes one degree under the current degree distribution. Then the probabilities of 0, 1 bits in output bits $c_{i}$ $(i\in {1,2,...,n})$ are given:

\begin{align}\label{Raptor_uniform_distribution}
&Prob(c_{i}=0) \nonumber \\
&=Prob(c_{i}=1) \nonumber \\
&=Prob(v_{g_{i},1}\bigoplus v_{g_{i},2}\bigoplus ...\bigoplus v_{g_{i},\omega_{i}}=0)\\ &=Prob(v_{g_{i},1}\bigoplus v_{g_{i},2}\bigoplus ...\bigoplus v_{g_{i},\omega_{i}}=1)\nonumber \\
&= \frac{1}{2}, \nonumber
\end{align}
where $Prob$ means probability. Thus the binary sequence $c$ generated by Raptor encoding is uniformly distributed. Binary sequence $c$ can not be directly used in the multidimensional reconciliation method, thus they need to be converted into binary spherical codes. Spherical codes mean that all code words lie on a sphere centered on 0.
Therefore a further conversion is needed such as the following:
\begin{equation}\label{further_conversion}
(c_{1},c_{2},...,c_{d})\rightarrow (\frac{(-1)^{c_{1}}}{\sqrt{d}},\frac{(-1)^{c_{2}}}{\sqrt{d}},...,\frac{(-1)^{c_{d}}}{\sqrt{d}}).
\end{equation}

After binary sequence $c$ is converted into binary spherical sequence $c'$, the mapping function $M(y',c')$ is calculated by Bob with sequence $y'$ and sent to Alice. The mapping function $M(y',c')$ satisfies:
\begin{equation}\label{cal_mapping_function}
M(y',c')\cdot y'=c'.
\end{equation}

The number of encoded sequences $c$ is required to be a multiple of $d$ because the dimension of multidimensional reconciliation is $d$. With the Raptor encoding, Bob calculates more mapping functions and sends them to Alice. Using the mapping functions, Alice can map her Gaussian variable $x'$ to $e$ where $e=M(y',c')\cdot x'$. Through the above steps, the reduction of the physical Gaussian channel is reformulated to a virtual BIAWGN channel.
Then Alice starts to recover $u$ by Raptor decoding.
$u'$ denotes the decoding result that is equals to $u$ when decoding is successful.
However, when the decoding is successful, there is also a certain probability that the decision is wrong. Hence it is necessary to send additional check codes $r$ for further judgment. If the decoding fails, Bob calculates more and more mapping functions, and Alice prepares for the next round of decoding.

In addition, the mapping functions $M(y',c')$ sent on a public channel do not give any information to Eve about $c'$. Here we consider the spherical code $C'=\left \{ c'_{1},c'_{2},...,c'_{N} \right \}$ and have $Prob(c'=c'_{i})=\frac{1}{N}$. According to the Haar measure, there exists a random orthogonal transformation $Q$ on $\mathbb{R}^{d}$, which maps $y'$ to random variable $y''$ on the mediator hyperplane, i. e. $Q(y')=y''$. Then $y''$ is converted to $c'$ by reflection transformation $P$, i. e. $c'=P(y'')$. The mapping function $M(y',c')=P\cdot Q$ is a random orthogonal transformation and satisfies \cite{leverrier2008multidimensional}:
\begin{equation}\label{security_analyze}
Prob(c'=c'_{i}\mid M(y',c'))=\frac{1}{N}.
\end{equation}
Therefore, $M(y',c')$ and $c'$ are independent and messages transmitted in the proposed rateless reconciliation protocol do not result in information leakage.

The rate of Raptor codes is uncertain before information transmission due to the property of rateless codes.
Let $n(\gamma )$ denote the number of coded bits required for the receiver to successfully decode the original $k$ bits using Raptor decoding at SNR $\gamma$. Then the realized rate of Raptor codes is defined as
\begin{equation}\label{realize_rate_of_Raptor_code}
R(\gamma)=k/\mathbb{E}[n(\gamma)],
\end{equation}
where $\mathbb{E}$ is the expectation operator, and $\mathbb{E}[n(\gamma)]$ denotes the average number of coded bits required for the successful decoding of the entire message bits.

Reconciliation efficiency is a significant parameter to evaluate the performance of the information reconciliation step. In this CV-QKD system, the efficiency of reconciliation is measured by
\begin{equation}\label{efficiency_reconciliation}
\beta (\gamma)=\frac{R(\gamma)}{C(\gamma)},
\end{equation}
where $R(\gamma)$ is the realized rate of Raptor codes, $C(\gamma)$ is the capacity of the quantum channel at SNR $\gamma$. In BIAWGN channel, $C(\gamma)$ is defined as
\begin{equation}\label{capacity_of_channel}
C(\gamma)=\frac{1}{2}\mathrm{log}(1+\gamma).
\end{equation}

\renewcommand\arraystretch{1.25}
\begin{table*}
	\caption{\label{tab:degree_distribution}The four main degree distributions in this work.} 
	\footnotesize\rm
	\begin{ruledtabular}
		\begin{tabular}{cm{16cm}<{} }
			Number & Degree distribution\\
			\hline
			$\Omega_{1}(x)$& $0.0269x^{320}+0.002x^{81}+0.022x^{60}+0.031x^{41}+0.0122x^{20}+0.0518x^{12}+0.0191x^{11}+0.0004x^{8}+0.0805x^{7}+0.0002x^{6}+0.0873x^{5}+0.0695x^{4}+0.2309x^{3}+0.3488x^{2}+0.0174x$ \\
			$\Omega_{2}(x)$& $0.027x^{300}+0.003x^{81}+0.022x^{60}+0.03x^{41}+0.0121x^{20}+0.0518x^{12}+0.0191x^{11}+0.0004x^{8}+0.0805x^{7}+0.0002x^{6}+0.0873x^{5}+0.0695x^{4}+0.2309x^{3}+0.3488x^{2}+0.0174x$ \\
			$\Omega_{3}(x)$& $0.0271x^{80}+0.002x^{61}+0.0218x^{40}+0.031x^{31}+0.0122x^{20}+0.0518x^{12}+0.0191x^{11}+0.0005x^{8}+0.0803x^{7}+0.0003x^{6}+0.087x^{5}+0.07x^{4}+0.2307x^{3}+0.3488x^{2}+0.0174x$ \\
			$\Omega_{4}(x)$& $0.0223x^{55}+0.0138x^{40}+0.0285x^{25}+0.0589x^{15}+0.0994x^{7}+0.0656x^{5}+0.0641x^{4}+0.1769x^{3}+0.4639x^{2}+0.0066x$ \\
		\end{tabular}
	\end{ruledtabular}
\end{table*}

It can be seen from Eq.~(\ref{finite_secret_key}) that the value of the reconciliaiton efficiency does affect the final secret key rate in CV-QKD system. When the quantum channel between Alice and Bob is stable, low efficiency leads to a low secret key rate.
Eqs.~(\ref{realize_rate_of_Raptor_code},\ref{efficiency_reconciliation}) show that $n(\gamma )$ is inversely proportional to the efficiency. If the secret key rate is less than 0, then we do not need to generate limitless bits to ensure decoding success. Therefore, it is necessary to limit the scope of $n(\gamma )$.
Let $\beta_{\mathrm{min}}$ denote the minimum reconciliation efficiency required for the system, which satisfies $K_{\mathrm{finite}}(\beta_{\mathrm{min}})$=0. Let $n_{\mathrm{val}}$ denote the maximum length of $n(\gamma )$ that meets the system requirements. If the number of the message bits needed for decoding is greater than $n_{\mathrm{val}}$, we abandon this original message $u$. Thus, $n(\gamma )$ satisfies the following condition:
\begin{equation}
\frac{k}{C(\gamma)}+ O(k')<n(\gamma )< n_{\mathrm{val}}=\frac{k}{\beta_{\mathrm{min}}\cdot C(\gamma)},
\end{equation}
where $k$ denotes the length of $u$ and $k'$ denotes the input bits number for the LT code. In addition, we can choose a fixed reconciliation efficiency, such as 96\%, to perform decoding once, which can support a high repetition frequency of the CV-QKD system.

In order to satisfy the requirement that the secure key rate is greater than zero, a higher reconciliation efficiency is needed under the condition of low SNRs. From Eq.~(\ref{efficiency_reconciliation}), we know that the realized rate $R(\gamma)$ is the major factor affecting efficiency. In other words, the efficiency of the proposed protocol depends to a great extent on the performance of the designed Raptor codes. The goal of Raptor-code design is to find the output node degree distribution to maximize the design rate of the LT code.
In this Paper, we obtain the degree distribution by the EXIT chart approach \cite{cheng2009design,shirvanimoghaddam2016raptor}, which is based on two assumptions. Firstly, all incoming messages arriving at a given node are statistically independent. Secondly, the degree of each input node is high, and the message sent from the node is approximately Gaussian.
Under the above two assumptions, the problem of finding the optimal degree distribution can be transformed into the problem of solving linear programming.
Four mainly degree distributions are used for Raptor codes in this work and their descriptions are shown in Table~\ref{tab:degree_distribution}.

\begin{figure}[b]
	\centering
	\includegraphics[width=1.08\linewidth]{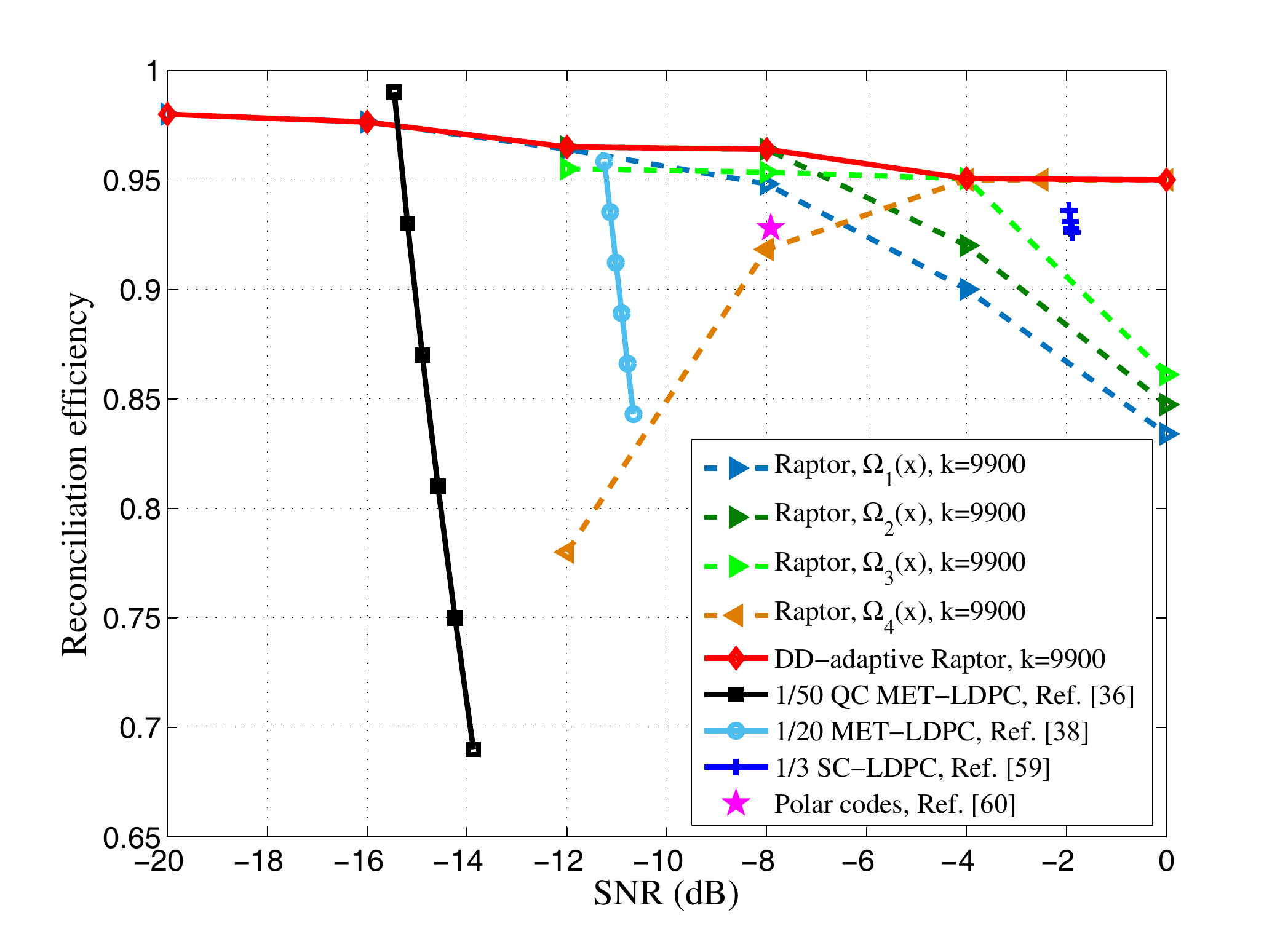}
	\caption{(Color online) Reconciliation efficiencies under different SNRs. The red line segment represents the efficiency based on the rateless reconciliation protocol in this Paper with degree distribution adaptive (DD-adaptive) method. From left to right, dotted lines represent efficiency performances based on degree distribution $\Omega_{1}(x)$, $\Omega_{2}(x)$, $\Omega_{3}(x)$, and $\Omega_{4}(x)$ respectively. Other works, including quasicyclic (QC) MET LDPC codes \cite{milicevic2018quasi}, MET LDPC codes \cite{jouguet2011long}, spatially coupled (SC) LDPC codes \cite{jiangxueqin2018SCLDPC} and polar codes \cite{jouguet2014high} are also shown here.}
	\label{fig:simulateresult}
\end{figure}
\section{Simulation results}
\label{sec4}

Figure~\ref{fig:simulateresult} shows the reconciliation efficiencies under different SNRs. We set the information block size as $k=9900$ bits and the LDPC code $V$ is a rate-0.99 LDPC code which is constructed the same way as in Ref.~\cite{cheng2009design}.
Dotted lines, from left to right, represent efficiency performances based on the degree distribution $\Omega_{1}(x)$, $\Omega_{2}(x)$, $\Omega_{3}(x)$, and $\Omega_{4}(x)$, respectively (see Table \ref{tab:degree_distribution} for details).
In Fig.~\ref{fig:simulateresult}, the reconciliation efficiency obtained by using $\Omega_{1}(x)$ decreases with the increase of SNR. And when SNR is higher than -12 dB, the efficiency is lower than that of using $\Omega_{2}(x)$. Therefore, the degree distribution adaptive method is used to automatically switch the degree distribution of Raptor codes with the change of SNR. This method is equivalent to using one degree distribution to keep the reconciliation efficiency high. In other words, the red line segment is an envelope that covers all the hightest reconciliation efficiency in different SNRs.
As can be seen in Fig.~\ref{fig:simulateresult}, the efficiencies in our work are larger than 95\% in the range of SNR from -20 to 0 dB. When the SNR is -20 dB, the efficiency reaches 98\%.
The efficiency of QC MET LDPC codes with rate 1/50 designed in \cite{milicevic2018quasi} and that of MET LDPC codes with rate 1/20 designed in \cite{jouguet2011long} dramatically drops as the SNR changes.
Therefore, Raptor codes with an optimized degree distribution can achieve more stable efficiencies in comparison with the fixed-rate LDPC codes under a wide range of SNRs.

\begin{figure}[b]
	\centering
	\includegraphics[width=1.0\linewidth]{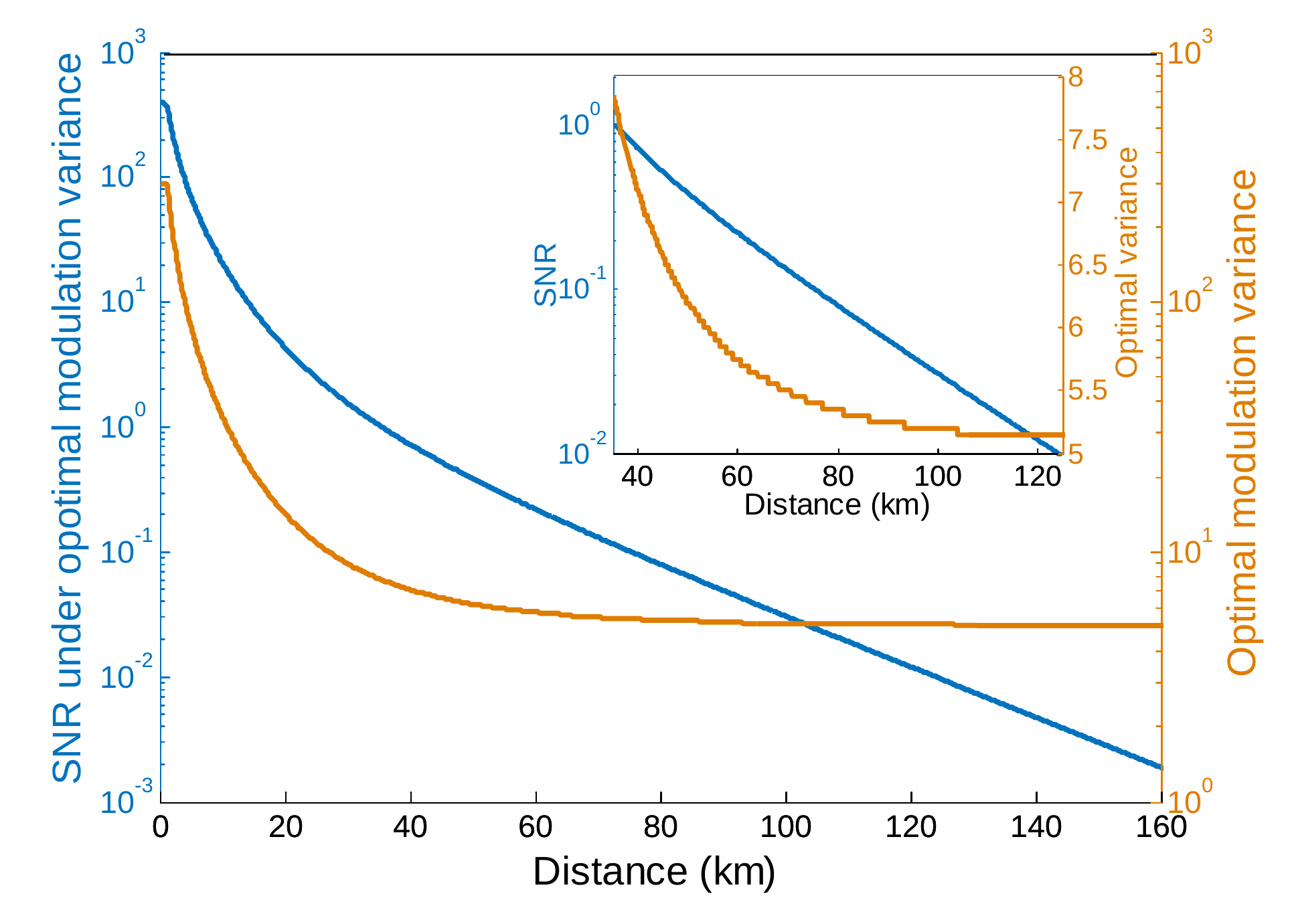}
	\caption{(Color online) Optimal modulation variance vs distance and the corresponding SNR. The top right-hand corner shows enlargement from 34 to 124 km. The parameters of our CV-QKD system are as follows: $\xi=0.01$ $\eta=0.6$, $\alpha=0.2dB/km$, $\beta=95.6\%$, and $\nu_{\mathrm{el}}=0.015$.}
	\label{fig:snr-distance}
\end{figure}

In previous CV-QKD systems, the modulation variance is usually adjusted in real time to get the code's target SNR. The main reason is to meet the threshold of available fixed-rate code and achieve high reconciliation efficiency.
The proposed protocol can achieve high reconciliation efficiency under a wide range of SNRs, so there is no need to adjust the modulation variance to meet the target SNRs. Here we can make it work at the optimal value to increase the secret key rate.
Figure~\ref{fig:snr-distance} shows the optimal modulation variance at the sender side with respect to the distance and the corresponding SNR. And it decreases with distance increasing and gradually stabilized after about 40 km. As the distance increases, the SNR decreases. Under this condition, the range of SNR from -20 dB to 0 dB corresponds to the distance changing from 35 to 124 km, which is indeed a wide range of distance.
\begin{figure}
	\centering
	\includegraphics[width=1.0\linewidth]{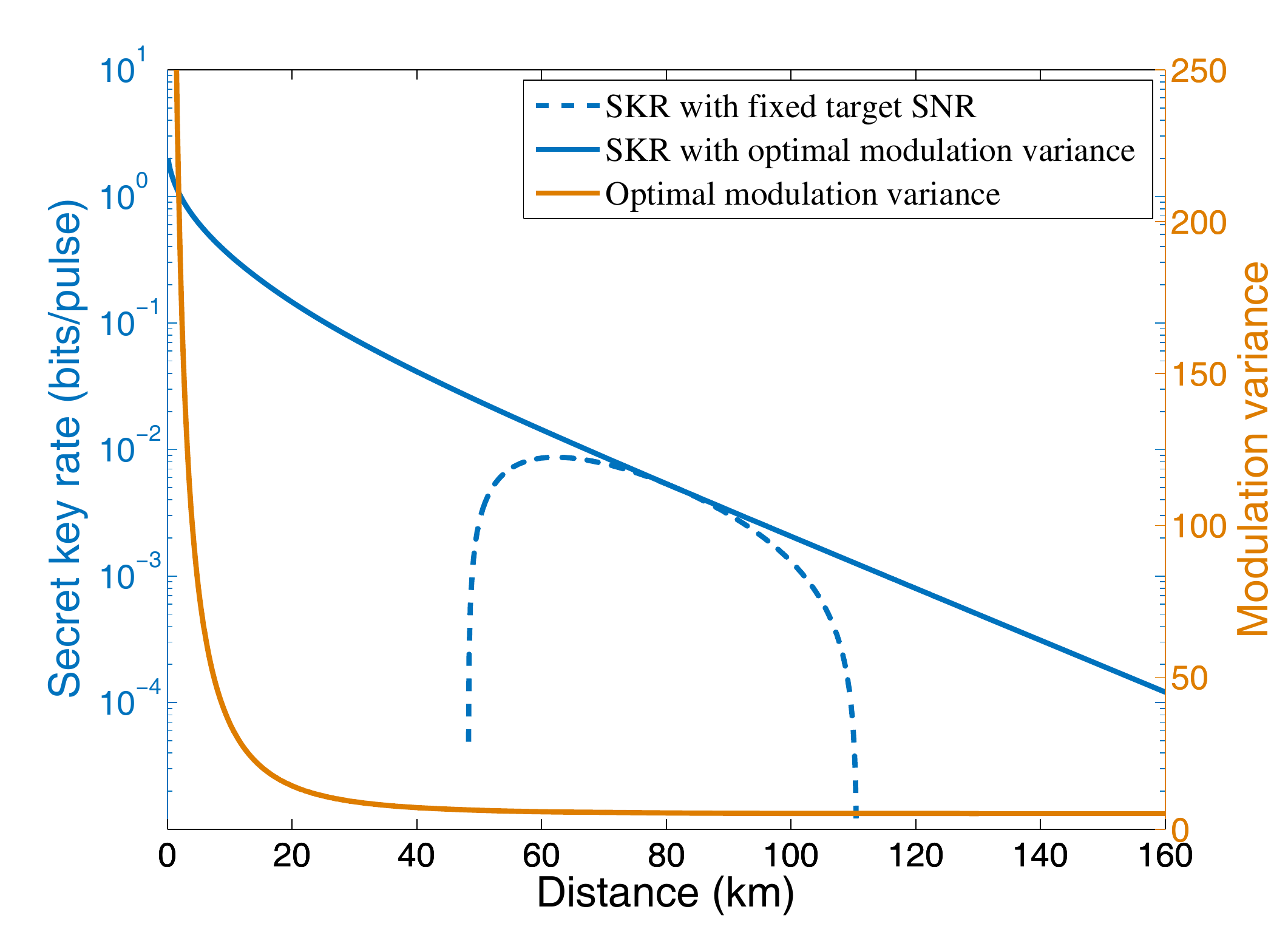}
	\caption{(Color online) Secret key rate with different modulation variances vs. distance. The blue solid line is the secret key rate for optimal modulation variance and the orange solid line is the corresponding optimal modulation variance. The blue dotted line is the secret key rate for the modulation variance adjusted in the real time according to the target SNR=0.075 and the orange dotted line is the corresponding modulation variance. Neither set of data considers system overhead. Other parameters are as follows: $\xi=0.01$ $\eta=0.6$, $\alpha=0.2dB/km$, $\beta=95.6\%$, and $\nu_{\mathrm{el}}=0.015$.}
	\label{fig:optimal-va}
\end{figure}
Figure~\ref{fig:optimal-va} shows the secret key rate under different modulation variances with respect to the distance.
When the modulation variance corresponding to the blue dotted line is consistent with the optimal modulation variance, there is one overlap of the solid line and the dotted line.
The blue dotted line indicates that the optimal modulation variance has to be sacrificed for high reconciliation efficiency of the fixed-rate code. Thus when the deviation between the modulation variance and the optimal value is large, the secret key rate will decreases rapidly.
Therefore, using optimal modulation variance can improve the performance of the CV-QKD system.

\begin{figure}
	\centering
	\includegraphics[width=1.0\linewidth]{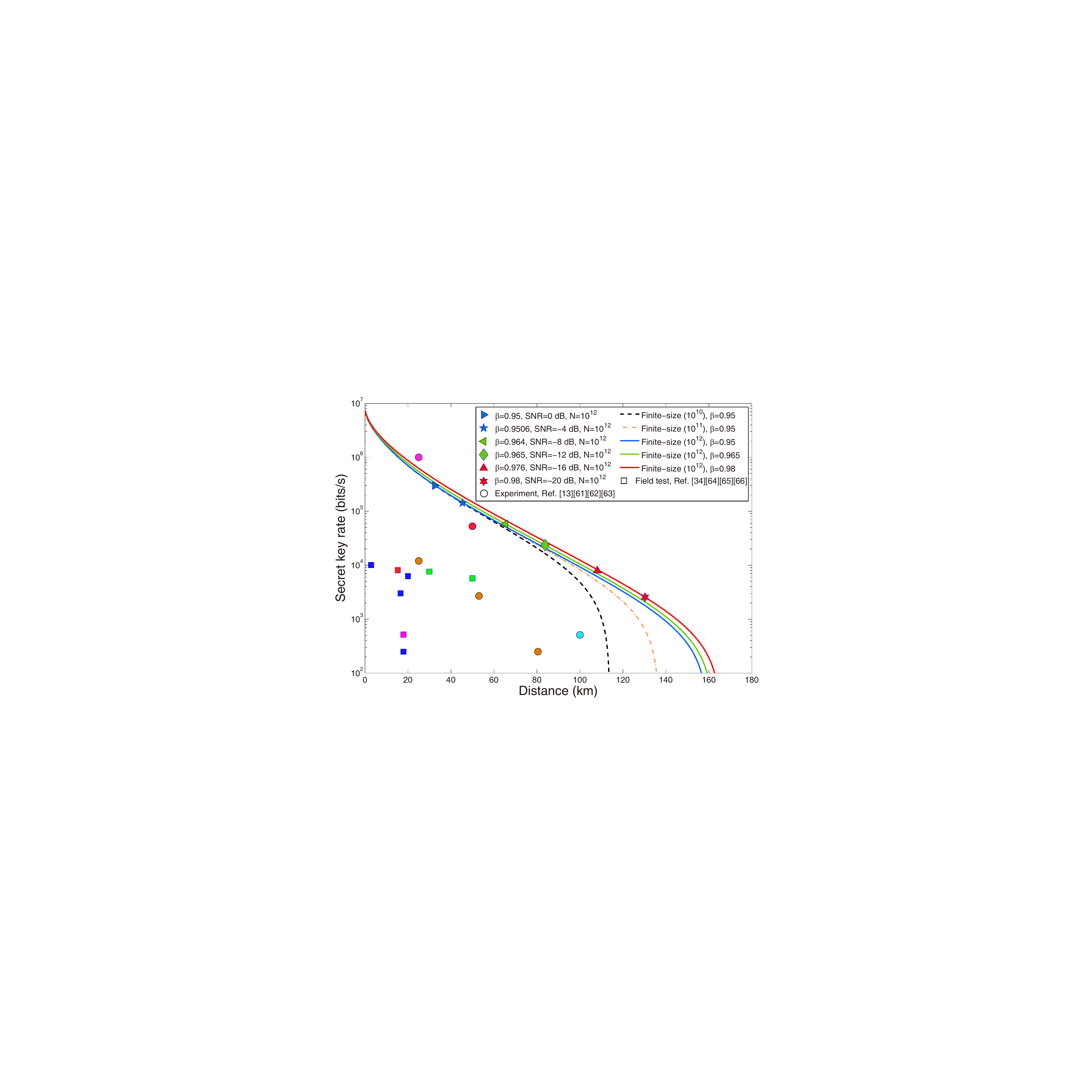}
	\caption{(Color online) Finite-size secret key rate with 5-MHz repetition rate vs. distance. The triangle points, five-pointed star, rhombus point, and six-pointed star correspond to our simulation results. Solid lines are asymptotic theoretical key rates. The first three lines from left to right correspond to block lengths of $N = 10^{10}, 10^{11}, 10^{12}$, respectively, and their reconciliation efficiency $\beta = 0.95$. The other solid lines represent the finite-size ($10^{12}$) theoretical key rates with different efficiencies. The modulation variance $V_{A}$ is always kept at the optimal value, $\beta$ refers to the result based on rateless reconciliation protocol in Fig.~\ref{fig:simulateresult}. The orange dots are the secret key rates obtained from the state-of-art experiment \cite{jouguet2013experimental,huang2016long,wang201525,Huang:15}. Green block points are the secret key rates obtained from the state-of-the-art fied test \cite{Zhang_2019,Huang:16,Jouguet:12,Fossier_2009}. Other parameters in our results are as follows: $\xi=0.01$ $\eta=0.6$, $\alpha=0.2dB/km$,  and $\nu_{\mathrm{el}}=0.015$.}
	\label{fig:skr}
\end{figure}

According to Eq.~(\ref{finite_secret_key}), the ratio of the data which is used to extract the secret keys to total data has an important impact on the secret key rate. In previous CV-QKD systems, almost half of the raw data is used for parameter estimation, which will reduce the secret key rate by 50\%. In our system, we swap the order of parameter estimation and information reconciliation. Thus, we can extract the keys from all the data to have an almost doubling of the final key rates \cite{Wang2019}.
Figure~\ref{fig:skr} shows the secret key rate with respect to the transmission distance.
The secret key rate we achieve here is based on the system with 5-MHz repetition rate.
In these simulation results, the rateless reconciliation protocol in Fig.~\ref{fig:reconciliation-protocol} is applied to improve the robustness of system and support high secret key rate. Furthermore, the modulation variance $V_{A}$ is adjusted in real time according to the system parameters to be as close as possible to the theoretical optimal value. However, it is unrealistic to obtain the results of all the values of SNRs in the range from -20 to 0 dB. Figure~\ref{fig:skr} shows the simulation results of block lengths $N = 10^{12}$ at SNR of -20, -16, -12, -8, -4 and 0 dB, respectively, and gives their asymptotic theoretical secret key rate. We also compare the key rate under different block lengths ($N = 10^{10}, 10^{11}$ and $10^{12}$).
Notably, when the block length is less than $10^{10}$, the secret key rate is less than zero at SNR=-20 dB.
In theory, highly efficient key extraction can be maintained at any SNR between -20 and 0 dB. In particular, the secret key rate is 300 kbit/s at 32 km (SNR=0 dB), and 2.5 kbit/s at 130 km (SNR=-20 dB). In Fig.~\ref{fig:skr}, the state-of-the-art experiment results, and field test results are given. It can be seen that the proposed protocol is comparatively advantageous.
Overall, our work provides a reference for the application of CV-QKD systems in different scenarios.

\section{Discussion}
\label{sec5}
The rateless reconciliation protocol is proposed in this Paper for CV-QKD that combines multidimensional reconciliation schemes and Raptor codes.
Compared with the fixed-rate code method, the proposed protocol has two outstanding features.
Firstly, the proposed protocol can achieve high reconciliation efficiency with just one degree distribution under a wide range of SNRs. It reduces the complexity of optimization and improves the robustness of the CV-QKD system.
Secondly, the modulation variance of the system is allowed to work at the optimal value which can significantly improve the secret key rate.

Extracting keys in long-distance CV-QKD systems depends heavily on highly efficient postprocessing at low SNR. In theory, the rateless reconciliation protocol can achieve error-correction under lower SNRs (-25 or -30 dB), which supports the practical application of long-distance CV-QKD systems. Another important future work is to improve the speed of the protocol.
In this Paper, the data processing is completed under offline conditions. The postprocessing in our work is completed on the CPU platform, thus the speed can not meet the real-time requirement of the system. In future works, the GPU platform can be used for parallel processing to improve the performance.

The rateless reconciliation protocol proposed in this work can maintain highly efficient key extraction under the wide range of SNRs and is suitable for CV-QKD systems in various scenarios. It also can be applied to two areas that show promise for QKD.
The first is the free-space QKD system that Alice sends quantum states to Bob without fiberoptic. Due to the fact that the transmittance fluctuation caused by atmospheric turbulence effects may introduce excess noise, the SNR varies greatly in a short time. This system needs high and stable reconciliation efficiency, and the proposed protocol can meet this requirement.
The second is the QKD network. Raptor codes are suitable for broadcasting systems, such as one-to-many star networks. Quantum states are sent from one sender to multiple receivers, and then secret key extraction is performed separately. The proposed protocol is helpful to realize quantum secret sharing in such networks.

\section{Acknowledgments}
This work was supported in part by the Key Program of National Natural Science Foundation of China under Grant No. 61531003, the National Natural Science Foundation under Grant No. 61427813, and the Fund of State Key Laboratory of Information Photonics and Optical Communications.

C. Z. and X. W. contributed equally to this work.

%


\end{document}